\documentclass[sigconf,natbib=true]{acmart}
\AtBeginDocument{%
  \providecommand\BibTeX{{%
    \normalfont B\kern-0.5em{\scshape i\kern-0.25em b}\kern-0.8em\TeX}}}

\setcopyright{acmlicensed}
\copyrightyear{2018}
\acmYear{2018}
\acmDOI{XXXXXXX.XXXXXXX}

\acmConference[SIGIR 2024]{The 47th International ACM SIGIR Conference on Research and Development in Information Retrieval}{July 14-18, 2024}{Washington D.C., USA}
\acmISBN{978-1-4503-XXXX-X/18/06}

\usepackage{multirow}
\usepackage{makecell}
\begin{document}

\title{Sequential Recommendation with Latent Relations based on Large Language Model}

\author{Shenghao Yang}
\affiliation{%
  \institution{DCST, Tsinghua University}
  \city{Beijing}
  \country{China}
  \postcode{100084}
}
\email{ysh21@mails.tsinghua.edu.cn}

\author{Weizhi Ma}
\authornote{Corresponding authors\\This work is supported by the Natural Science Foundation of China (Grant No. U21B2026, 62372260) and Quan Cheng Laboratory (Grant No. QCLZD202301).
}
\affiliation{%
  \institution{AIR, Tsinghua University}
  \city{Beijing}
  \country{China}
  \postcode{100084}
}
\email{mawz@tsinghua.edu.cn	}

\author{Peijie Sun}
\affiliation{%
  \institution{DCST, Tsinghua University}
  \city{Beijing}
  \country{China}
  \postcode{100084}
}
\email{sun.hfut@gmail.com}

\author{Qingyao	Ai}
\affiliation{%
  \institution{DCST, Tsinghua University}
  \city{Beijing}
  \country{China}
  \postcode{100084}
}
\email{aiqy@tsinghua.edu.cn}

\author{Yiqun	Liu}
\affiliation{%
  \institution{DCST, Tsinghua University}
  \city{Beijing}
  \country{China}
  \postcode{100084}
}
\email{yiqunliu@tsinghua.edu.cn}

\author{Mingchen Cai}
\affiliation{%
  \institution{Meituan}
  \city{Beijing}
  \country{China}
}
\email{caimingchen@meituan.com}

\author{Min	Zhang}
\authornotemark[1]
\affiliation{%
  \institution{DCST, Tsinghua University}
  \city{Beijing}
  \country{China}
  \postcode{100084}
}
\email{z-m@tsinghua.edu.cn}

\renewcommand{\shortauthors}{Yang, et al.}

\begin{abstract}
Sequential recommender systems predict items that may interest users by modeling their preferences based on historical interactions. Traditional sequential recommendation methods rely on capturing implicit collaborative filtering signals among items. 
Recent relation-aware sequential recommendation models have achieved promising performance by explicitly incorporating item relations into the modeling of user historical sequences, where most relations are extracted from knowledge graphs. However, existing methods rely on manually predefined relations and suffer the sparsity issue, limiting the generalization ability in diverse scenarios with varied item relations.

In this paper, we propose a novel relation-aware sequential recommendation framework with \textbf{L}atent \textbf{R}elation \textbf{D}iscovery (LRD). Different from previous relation-aware models that rely on predefined rules, we propose to leverage the Large Language Model (LLM) to provide new types of relations and connections between items. The motivation is that LLM contains abundant world knowledge, which can be adopted to mine latent relations of items for recommendation. 
Specifically, inspired by that humans can describe relations between items using natural language, LRD harnesses the LLM that has demonstrated human-like knowledge to obtain language knowledge representations of items. These representations are fed into a latent relation discovery module based on the discrete state variational autoencoder (DVAE). Then the self-supervised relation discovery tasks and recommendation tasks are jointly optimized. 
Experimental results on multiple public datasets demonstrate our proposed latent relations discovery method can be incorporated with existing relation-aware sequential recommendation models and significantly improve the performance. Further analysis experiments indicate the effectiveness and reliability of the discovered latent relations.
\end{abstract}



\keywords{Sequential recommendation, Large language model, Latent relation}



\maketitle

\section{Introduction}
Sequential recommendation is a promising researched topic in the community of recommender systems, aiming to predict the next item the user prefers based on his/her interaction history~\cite{wang2019sequential}. Various methods have been proposed to address the sequential recommendation task. Early studies focused on estimating the transition relations between items based on Markov chain assumptions~\cite{rendle2010factorizing}. In recent years, with the advancement of deep learning, various deep neural networks, such as Recurrent Neural Networks (RNNs)~\cite{hidasi2015session,li2017neural,jang2020cities}, Convolutional Neural Networks (CNNs)~\cite{tang2018personalized}, and Transformers~\cite{kang2018self,10027680,he2021locker,sun2019bert4rec}, have been incorporated to better model user preferences reflected in their sequential historical interactions.

Although existing methods have achieved remarkable performance, they usually rely on item-based collaborative filtering algorithms~\cite{sarwar2001item} to calculate the implicit collaborative similarity between items while overlooking the explicit relations between items, which are prevalent and significantly influence user decisions in the real recommendation scenario.
Recently, some relation-aware sequential recommendation methods have been proposed~\cite{wang2020make,xin2019relational,wang2020toward} to explicitly consider item relations during modeling user preferences and significantly improve the performance of the sequential recommendation. However, current approaches still face some challenges that limit the application of these models.

Specifically, for most existing relation-aware methods, item relation data is typically stored in a knowledge graph, which may suffer from \textit{sparsity} issue of two aspects. Firstly, the \textit{relation sparsity} on the edge set. The relational item modeling of existing methods is performed based on manually predefined relations. The attribute-based relations (e.g., ``share category'') and co-occurrence-based relations (e.g., ``also buy'') are usually used as these relations are relatively straightforward to define and can be derived from user interaction data and item metadata. Nevertheless, the relations between items are diverse in the real world, and manually defined relations are sparse compared to all latent relations. Relying on a restricted set of predefined relations limits the model's capacity to generalize effectively across diverse recommendation scenarios. Secondly, the \textit{item sparsity} on the vertex set. It is caused by the inherent data sparsity issue of recommender systems~\cite{roy2022systematic} and particularly affects the data collection of the co-occurrence-based relation since it requires substantial interaction data to collect item pairs that conform to the relation definition.
To alleviate the above issue, we investigate to discover latent relations between items that contribute to the recommendation.

In this paper, we propose a language knowledge-based \textbf{L}atent \textbf{R}elation \textbf{D}iscovery (LRD) method for the sequential recommendation. The motivation behind this approach is inspired by the fact that humans usually describe relations between items in natural language based on their knowledge. Observing the rich world knowledge and semantic representation capabilities exhibited by Large Language Models (LLMs)~\cite{touvron2023llama,brown2020language,openai2023gpt4}, we propose to leverage the abilities of LLMs to discover latent item relations. 
Specifically, we design a self-supervised learning framework to facilitate the process of discovering latent relations. 
We first leverage an LLM to obtain language knowledge representations of items.
Subsequently, a relation extraction model is adopted to predict the latent relation between two items. 
Then we incorporate an item reconstruction model to reconstruct the representation of one item based on the representation of the predicted relation and the other item.
Through this self-supervised learning process, the objective of reconstructing the original items forces the relation extraction model to predict relations with sufficient accuracy and generality.
Furthermore, we incorporate the LRD into the existing relation-aware sequential recommendation frameworks and perform joint optimization.

The merits of our proposed framework are threefold. 
Firstly, LRD does not rely on manually defined relations, and can autonomously discover latent relations between items. This enhances the model's ability to better capture diverse preferences reflected in user interaction history and improve recommendation performance. 
Secondly, the optimal objective of the relation-aware sequential recommendation task serves as supervised signals to guide the relation discovery process, leading to the discovery of relations more beneficial to the recommendation. 
Last but not least, analyzing the predicted item relations by the LRD contributes to better interpretability of relation-aware sequential recommendation models.

We perform experiments on multiple public datasets to evaluate our proposed LRD approach. Leveraging latent relations derived from language knowledge-based item representation, the relation-aware sequential recommendation model captures more comprehensive user sequence representations, significantly improving the performance of sequential recommendation. Experimental results demonstrate that compared to state-of-the-art (SOTA) relation-aware sequential recommendation models, the model enhanced by LRD achieves significantly better performance. Further analysis experiments reveal that the LRD module is indeed capable of discovering reasonable relations between items.

The main contributions of our work are summarized as follows:
\begin{itemize}
\item To the best of our knowledge, we first propose to discover latent relations based on LLM for relation-aware sequential recommender systems.
\item We propose an LLM-based latent relation discovery framework, i.e., LRD, to harness the language knowledge to discover latent relations, which is a self-supervision learning method and flexible to work with existing relation-aware sequential recommenders through joint learning. 
\item Experimental results on multiple public datasets demonstrate that LRD significantly improves the performance of existing relation-aware sequential recommendation models by effectively discovering reliable relations between items.
\end{itemize}

\begin{figure*}[t] 
  \centering
  \includegraphics[width=\linewidth]{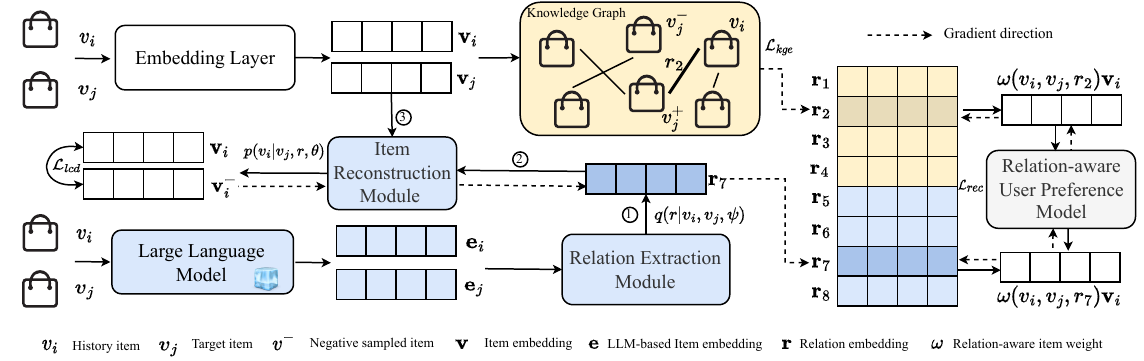}
  \caption{Overall Framework of relation-aware sequential recommendation with LRD. There are two main components of LRD: a relation extraction model to estimate the latent relation based on the language knowledge representation of two items obtained by LLM and an item reconstruction model to reconstruct the item based on the estimated relation and another item. The predefined relations from the knowledge graph and latent relations from LRD are both used to contribute to the relational item modeling in the recommendation.}
\label{fig: lrd}
\end{figure*}

\begin{table}[t]
\tabcolsep=2pt
\begin{center}
\caption{Notations.}
  \label{tab: nota}
\begin{tabular}{ll}
\toprule
Notations & Descriptions \cr
\midrule
$\mathcal{U}$ & The set of user \cr
$\mathcal{V}$ & The set of item \cr
$\mathcal{T}$ & The set of triplets \cr
$\mathcal{R}$ & The set of relations\cr
$\mathcal{R}_{def}$ & The set of predefined relations\cr
$\mathcal{R}_{latent}$ & The set of latent relations\cr
$S_u$ & the interaction sequence of user $u$ \cr
$\mathbf{u} \in \mathbb{R}^{d}$ & The id embedding of user $u$ \cr
$\mathbf{m}_{u,v} \in \mathbb{R}^{d}$ & The relation-aware user sequence representation of $u$ \cr
$\mathbf{v} \in \mathbb{R}^{d}$ & The id embedding of item $v$ \cr
$\mathbf{e} \in \mathbb{R}^{d_{L}}$ & The LLM-based embedding of item $v$ \cr
$\mathbf{r} \in \mathbb{R}^{d}$ & The embedding of relation $r$ \cr
\bottomrule
\end{tabular}
\end{center}
\end{table}

\section{Problem Statement}
Firstly, the notations used in this paper are defined in Table~\ref{tab: nota} with descriptions.

Let $\mathcal{U}$ and $\mathcal{V}$ denote the sets of users and items, respectively. For each user $u \in \mathcal{U}$, its chronologically-ordered interaction history is represented as $S_u = \{v_1, v_2, v_3, ..., v_{N_u}\}$. For an item $v_i \in \mathcal{V}$, there may exist another related item $v_{-i}$ with relation $r$, denoted as a triplet $(v_i, v_{-i}, r)$. 
$\mathcal{R}$ denotes the set of relations, which is further divided into predefined relations set $\mathcal{R}_{def}$ and latent relations set $\mathcal{R}_{latent}$. 
All relational item triplets associated with predefined relations $\mathcal{T}$ can be stored in a knowledge graph $\mathcal{G}$, where the vertex set comprises all relational item pairs, and the edge set consists of all the predefined relations.

The objective of the sequential recommendation task is to provide a ranked list of items for the user $u$ at the next interaction, considering their interaction history $S_u$. The relation-aware sequential recommendation further considers the relation between each historic item $v_i\in S_u$ and the target item $v_j$. The existing methods only consider the relations in $R_{def}$, while our method LRD further incorporates a latent relations set $R_{latent}$. 

\section{Method}
\subsection{Framework Overview}
The framework of our proposed relation-aware sequential recommendation based on latent relation discovery is illustrated in Figure~\ref{fig: lrd}. 

The key component of the framework is the latent relation discovery module, which is designed as a self-supervised learning process inspired by the Discrete-state Variational Autoencoder (DVAE)~\cite{marcheggiani2016discrete}.
The latent relation discovery model comprises two submodules: 1) A relation extraction module that utilizes an LLM to obtain language knowledge representations of items and predicts the latent relation between two items based on their language knowledge representations. 2) An item reconstruction module that reconstructs the representation of one item based on the representation of the predicted latent relation and the other item.
Here, we incorporate the latent relation discovery module into the relation-aware sequential recommender. 
Specifically, we use the predicted latent relations to extend the predefined item relation embedding to better construct the user preference model. At the same time, we use the objectives of recommendation tasks to guide the discovery of more useful relations.

Next, we introduce the overall design of the latent relation discovery approach in Section~\ref{sec: lrd}, followed by the pipeline of the relation-aware sequential recommendation model based on latent relation discovery in Section~\ref{sec: rel_aware_SR}.

\subsection{Latent Relation Discovery (LRD)} \label{sec: lrd}

\subsubsection{Optimization Objective} \label{sec: learning_obj}
Our objective is to predict the latent relation between two items. Since latent relations are those not covered by manually crafted relation datasets, we cannot train the model with supervised learning. Instead, we adopt a self-supervised learning method inspired by DVAE. Following~\cite{marcheggiani2016discrete}, we assume that all relations follow a uniform distribution $p_u(r)$, and the optimization objective is formalized as the following pseudo-likelihood:
\begin{equation}
\begin{aligned}
\label{equ: psdlld}
\mathcal{L}(\theta) &= \log \sum_{r\in\mathcal{R}}p(v_i,v_{-i}|r,\theta)p_u(r) \\
&\approx \sum_{i=1}^2 \log \sum_{r\in\mathcal{R}}p(v_{i}|v_{-i},r,\theta)p_u(r),
\end{aligned}
\end{equation}
where $v_i$ and $v_{-i}$ denote a pair of items, $i \in \{1,2\}$. $v_{-i}$ denotes $\{v_1,v_2\}\backslash\{v_i\}$ and $p(v_{i}|v_{-i},r,\theta)$ denotes the probability of one item given another item and a relation, where $\theta$ denotes the parameters set.

The Pseudo-likelihood $\mathcal{L}(\theta)$ can be lower-bounded based on Jensen's inequality through a variational posterior $q(r|v_i,v_{-i},\psi)$:
\begin{equation}
\label{equ: lrd_obj}
\begin{aligned}
\mathcal{L}(\theta) > \mathcal{L}(\theta,\psi) = &\sum_{i=1}^2 \sum_{r\in\mathcal{R}}q(r|v_i,v_{-i},\psi) \log p(v_{i}|v_{-i},r,\theta) \\ &+ \alpha H[q(r|v_i,v_{-i},\psi)],
\end{aligned}
\end{equation}
where $q(r|v_i,v_{-i},\psi)$ is the relation extraction model used to predict the relation between a pair of items and $p(v_{i}|v_{-i},r,\theta)$ is the item reconstruction model that reconstructs the representation of the item given the predicted relation and another item. $\psi$ and $\theta$ denote the parameters of the two models, respectively. Intuitively, maximizing the probability of reconstructing the original item force $q(r|v_i,v_{-i},\psi)$ to provide sufficiently accurate and highly generalizable relations. $H$ is an entropy term used to regularize the probabilities predicted by the relation extraction model, ensuring more uniform predictions. $\alpha$ is the hyper-parameter to balance the regularization strength. Next, we further present the design of the relation extraction model $q(r|v_i,v_{-i},\psi)$ and the item reconstruction model $p(v_{i}|v_{-i},r,\theta)$ separately.

\subsubsection{Relation Extraction} \label{sec: rel_ex}
In the relation extraction model, we aim to predict the latent relation between two given items. Typically, well-defined relations between items can be conveniently annotated and collected. For example, attribute-based relations only require the collection of items with the same attribute. Nevertheless, in real-world scenarios, relations between items are more complex and diverse, making them challenging to define manually. Consequently, relying solely on predefined relations has limitations, as it cannot adequately model user preferences reflected in the interaction history. Intuitively inspired by the ability of humans to describe relations between two items with natural language based on their knowledge, we investigate discovering latent relations between items from the perspective of language knowledge. Considering that LLM has exhibited human-like world knowledge and effective semantic representations, we leverage LLM to extract language knowledge representations of items and feed them into the relation extraction model.

Specifically, given an item $v = \{w_1, w_2, w_3, ..., w_{N_v}\}$, where $w_i$ denotes each token of the item's text. We feed the token sequence into the LLM to obtain the language knowledge representation of the item, as shown in Equation~(\ref{equ: llm}).
\begin{equation} 
\label{equ: llm}
\begin{aligned}
\mathbf{e} = W_1(LLM([w_1, w_2, w_3, ..., w_{N_i}])) + b_1,
\end{aligned} 
\end{equation}
where $LLM(\cdot)$ denotes a specific pooling strategy on the last hidden state of the LLM to obtain the output item representation. Different LLMs may use different pooling strategies, such as CLS-pooling, mean-pooling, etc~\cite{li2020sentence}. $W_1 \in \mathbb{R}^{d_{L} \times d}$ and $b_1 \in \mathbb{R}^{d}$ denote the weight and bias of a projection layer, respectively, which is used to reduce the dimensionality of the LLM's output to match the input dimensions of the recommendation model.

With the enriched world knowledge of the LLM, we obtain item representations that potentially embed important information for the discovery of item relations not covered in the manually predefined relation set, which we refer to as the language knowledge representations. Next, the relation extraction model $q(r|v_i,v_{-i},\psi)$ in Equation~(\ref{equ: lrd_obj}) predicts the relation between two given items on the relation set $\mathcal{R}$ based on their language knowledge representations, i.e., $\mathbf{e}_i\in \mathbb{R}^d$ and $\mathbf{e}_{-i} \in \mathbb{R}^d$. In practice, we can adopt any classifier that allows gradient backpropagation. Without loss of generality, we use a lightweight linear classifier:
\begin{equation} 
\begin{aligned}
q(r|v_i,v_{-i},\psi) = SoftMax(W_2 [\mathbf{e}_i; \mathbf{e}_{-i}] + b_2),
\end{aligned} 
\end{equation}
where $W_2 \in \mathbb{R}^{2d \times |\mathcal{R}|}$ and $b_2 \in \mathbb{R}^{|\mathcal{R}|}$ are the weight and bias of the linear classifier, respectively, and $[;]$ denotes the concatenation operation.

\subsubsection{Relational Item Reconstruction} \label{sec: rel_item_recons}
Through the relation extraction model, we estimate the latent relation between two items $v_i$ and $v_{-i}$. Given the estimated relation and one of the items, the item reconstruction model aims to reconstruct the other item. The specific definition is as follows:
\begin{equation}\label{equ: item_recons}
\begin{aligned}
p(v_i|v_{-i}, r, \theta) = \frac{\exp(\phi(v_i, v_{-i}, r))}{\sum_{v_i' \in \mathcal{V}}\exp(\phi(v_i', v_{-i}, r))},
\end{aligned} 
\end{equation}
where $\phi(v_i, v_{-i}, r, \theta)$ is a scoring function for two items and a relation, and any triplet scoring function can be used. Without loss of generality, we use DistMult~\cite{yang2015embedding} as scoring functions:
\begin{equation}  \label{equ: tri_score}
\begin{aligned}
\phi(v_i, v_{-i}, r) = \mathbf{v}_i^T \text{diag}(\mathbf{r}) \mathbf{v}_{-i},
\end{aligned} 
\end{equation}
where $\text{diag}(\mathbf{r})$ is a diagonal matrix with the relation embedding $\mathbf{r}$ as its diagonal elements. Note that we use the representation of item IDs $\mathbf{v}_i$ for reconstruction instead of language knowledge representation $\mathbf{e}_i$. This design is to align the representation space of the predicted relation $r$ and the relation-aware recommendation model.

Since Equation~(\ref{equ: item_recons}) involves calculations across the entire set of items $\mathcal{V}$, resulting in high computational complexity, we approximate $\log p(v_i|v_{-i}, r, \theta)$ using negative sampling as follows:
\begin{equation}
\begin{aligned}
\log p(v_i|v_{-i}, r, \theta) =& \log \sigma(\phi(v_i, v_{-i}, r, \theta) \\
&+ \log \sigma(-\phi(v_i^-, v_{-i}, r, \theta))),
\end{aligned}
\end{equation}
where $\sigma$ is the sigmoid activation function, and $v_i^-$ is a randomly sampled negative item. Ultimately, the optimization objective in Equation~(\ref{equ: lrd_obj}) becomes:
\begin{equation}
\label{equ: final_lrd_obj}
\begin{aligned}
\mathcal{L}(\theta, \psi) = &\sum_{i=1}^2 \sum_{r \in \mathcal{R}} q(r|v_i, v_{-i}, \psi) [\log \sigma(\phi(v_i, v_{-i}, r, \theta) \\ 
&+ \log \sigma(-\phi(v_i^-, v_{-i}, r, \theta)))] + \alpha H[q(r|v_i, v_{-i}, \psi)].
\end{aligned}
\end{equation}

\subsection{LRD-based Sequential Recommendation} \label{sec: rel_aware_SR}
In this section, we present how to apply latent relations extracted by LRD into a relation-aware sequential recommender. We first introduce the pipeline of the sequential recommendation explicitly considering item relations in Section~\ref{sec: rel_sr}, followed by the joint optimization framework of the latent relation discovery and relation-aware sequential recommendation in Section~\ref{sec: joint_learn}.

\subsubsection{Relation-aware Sequential Recommendation} \label{sec: rel_sr}
Given a user's interaction history $S_u = \{v_1, v_2, v_3, ..., v_{N_u}\}$ and a target item $v_j$, the user's preference can be reflected in the history of interacted items. We further explicitly consider the relations between each historical item and the target item for sufficient user preference modeling. Formally, the preference score of user $u$ for target item $v_j$ is defined as:
\begin{equation} \begin{aligned}
y_{u, j} = (\mathbf{u} + \mathbf{m}_{u, j}) \mathbf{v}_j^T + b_j,
\end{aligned} \end{equation}
where $\mathbf{u} \in \mathbb{R}^{d}$ and $\mathbf{v}_j\in \mathbb{R}^d$ are representations of the user and target item, $\mathbf{m}_{u, j}\in \mathbb{R}^d$ is the user's historical sequence representation considering the relations between historical items and the target item. $b_j \in \mathbb{R}^d$ is the bias term.

Calculating $\mathbf{m}_{u, j}$ is a crucial step in modeling relation-aware user preferences. Specifically, it is the aggregation of multiple user sequence representations considering different relations types, which is defined as:
\begin{equation} \begin{aligned}
\mathbf{m}_{u, j} = \text{AGG}([\mathbf{s}_{u_j, r_1}; \mathbf{s}_{u_j, r_2}; ...; \mathbf{s}_{u_j, r_{|\mathcal{R}|}}]),
\end{aligned} \end{equation}
where \text{AGG} is an aggregation function that can adopt various aggregation methods such as mean-pooling, max-pooling, and attention-pooling~\cite{xin2019relational,wang2020toward}. $\mathcal{R}$ is the relations set, including predefined relations and latent relations discovered by the LRD. $\mathbf{s}_{u_j, r}$ is the historical sequence representation of user $u$ given a relation $r$ and target item $v_j$, defined as:
\begin{equation}
\begin{aligned}
\mathbf{s}_{u_j, r} = \sum_{v_i \in S_u} \omega(v_i, v_j, r) \mathbf{v}_i,
\end{aligned} 
\end{equation}
where $w(v_i, v_j, r)$ is the relation intensity between historical item $v_i$ and target item $v_j$ under relation $r$. It is a normalized weight across all relations, defined as:
\begin{equation} \begin{aligned}
\omega(v_i, v_j, r) = \frac{\exp(\phi(v_i, v_j, r))}{\sum_{v_i' \in \mathcal{V}/S_u}\exp(\phi(v_i', v_j, r))},
\end{aligned} \end{equation}
where $\phi(v_i, v_j, r)$ is the triplet scoring function given two items and a relation. The scoring function used here is consistent with the one used in the item reconstruction model in Section~\ref{sec: rel_item_recons}, i.e., Equation~(\ref{equ: item_recons}). This alignment facilitates joint optimization in subsequent steps.

\subsubsection{Joint Learning} \label{sec: joint_learn}
For the relation-aware sequential recommendation task, we adopt the BPR pairwise loss~\cite{rendle2012bpr} to define its optimization objective:
\begin{equation} \begin{aligned} \label{equ: rec_obj}
\mathcal{L}_{rec} = -\sum_{u \in \mathcal{U}}\sum_{j=2}^{N_u}\log \sigma (y_{u,j} - y_{u,j^-}).
\end{aligned} \end{equation}

To leverage discovered latent relations for the recommendation task and simultaneously let user interaction data guide the relation discovery process, we jointly optimize the objectives of the latent relation discovery task in Equation~(\ref{equ: final_lrd_obj}) and the recommendation task in Equation~(\ref{equ: rec_obj}). For this purpose, we make little modifications to the Equation~(\ref{equ: final_lrd_obj}) and explicitly represent item pairs as historical items and target items, as shown below:
\begin{equation} \begin{aligned}
\mathcal{L}_{lrd} = -\sum_{u \in \mathcal{U}}\sum_{j=2}^{N_u}\sum_{i=1}^{j-1}\sum_{r \in \mathcal{R}} & q(r|v_i,v_{j},\psi) [\log \sigma(\phi(v_i,v_{j},r,\theta)\\
&+ \log \sigma(-\phi(v_i^{-},v_{j},r,\theta))] \\
&+ \alpha H[q(r|v_i,v_j,\psi)].
\end{aligned} \end{equation}

Additionally, to ensure the model's capability to model predefined relations. We organize the set of triplets $\mathcal{T}$ with predefined relations into a knowledge graph and adopt a widely used knowledge graph embedding method to optimize the representations of items and relations in the knowledge graph, and the optimization objective is shown as:
\begin{equation} \begin{aligned}
\mathcal{L}_{kge} = -\sum_{(v_i,v_{-i},r) \in \mathcal{T}} \log \sigma (\phi(v_i,v_{-i},r) - \phi(v_i^-,v_{-i}^-,r)).
\end{aligned} \end{equation}

The knowledge graph embedding task is also included in the joint optimization framework. Ultimately, the joint optimization objective is defined as:
\begin{equation} \begin{aligned}
\mathcal{L} = \mathcal{L}_{rec} + \gamma \mathcal{L}_{kge} + \lambda \mathcal{L}_{lrd},
\end{aligned} \end{equation}
where $\gamma$ and $\lambda$ are the coefficients of the knowledge graph embedding task and the latent relation discovery task, respectively.

\section{Experiments}

In this section, we first introduce the experimental setting and then
present experimental results and analyses to answer the following research questions:
\begin{itemize}
    \item \textbf{RQ1}: How is the effectiveness of LRD-enhanced relation-aware sequential recommendation models?
    \item \textbf{RQ2}: Does each component of LRD-enhanced relation-aware sequential recommender contribute to the recommendation performance?
    \item \textbf{RQ3}: Does LRD discover reliable and vital relations between items?
\end{itemize}

\subsection{Experimental Settings}
\subsubsection{Datasets}
We conduct experiments on three publicly available datasets from diverse domains to validate the model's capability to discover latent relations in various recommendation scenarios.

\begin{itemize}
    \item \textbf{MovieLens}\footnote{https://grouplens.org/datasets/movielens/100k/}: This dataset is widely used for movie recommendations and consists of user ratings and attribute information for movies. We utilized the MovieLens-100k version for our experiments. Two predefined relations, namely ``release year'' and ``genre'', are extracted from the dataset. To discover latent relations between items, in addition to the movie title, release year, and genre available in the dataset, we crawl movie information from IMDB\footnote{https://www.imdb.com/}, including director, actor, and brief based on movie titles and release year.
    \item \textbf{Amazon Office Products and Electronics}~\cite{He_2016,mcauley2015imagebased}: These two datasets are subsets from the Amazon e-commerce dataset, representing two distinct domains. They contain user ratings, reviews, and rich item metadata. We utilize two attributes (i.e., category and brand) and co-occurrence information (i.e., ``also buy” and ``also view") as predefined item relations. Texts of the item title, category, and brand are used to discover latent relations between items.
\end{itemize} 

For all datasets, following previous work~\cite{wang2020toward}, we filter users and items with fewer than 5 interactions. The statistical information for each dataset after preprocessing is presented in Table~\ref{tab: dataset}.

\begin{table}[t]
\tabcolsep=3pt
\caption{Statistics of the datasets after preprocessing.}
\label{tab:datasets}
\begin{tabular}{clccc}
\toprule
 &\textbf{Datasets} & \textbf{MovieLens} & \textbf{Offices} & \textbf{Electronics}  \\
\midrule
\multirow{3}{*}{\makecell{User-Item\\Interactions}} &\#user & 943 & 4,905 &  192,403\\
    &\#item & 1,349 & 2,420 &  63,001\\
    &\#inter. & 99,287 & 53,258 &  1,682,498\\
    &density & 7.805\% & 0.448\% & 0.014\% \\
\midrule
\multirow{2}{*}{\makecell{Item\\Relations}} &\#relation & 2 & 4 & 4\\
    &\#triplets& 886K & 778K & 2,148M \\
\bottomrule
\end{tabular}
\label{tab: dataset}
\end{table}

\begin{table*}[t]
\tabcolsep=4pt
\caption{Overall performance of different models. The best performances are denoted in bold fonts. ``H@K” is short for ``HR@K” and ``N@K” is short for ``NDCG@K”, respectively. The subscript ``LRD” denotes the model is enhanced by LRD. “Improv.” means the relative improvement of the LRD-based model over the corresponding vanilla model. The superscripts $^{\dagger}$ and $^{\ddagger}$ indicate $p\leq0.05$ and $p\leq0.01$ for the paired t-test of the LRD-based model vs. vanilla model.}
\label{tab: overall}
\begin{tabular}{lllllllllllll}
\toprule
Datasets & \multicolumn{4}{c}{MovieLens}      & \multicolumn{4}{c}{Office}         & \multicolumn{4}{c}{Electronics}     \\
\cmidrule(lr){1-1} \cmidrule(lr){2-5} \cmidrule(lr){6-9} \cmidrule(lr){10-13}
Metrics  & H@5   & H@10  & N@5 & N@10 & H@5   & H@10  & N@5 & N@10 & H@5   & H@10  & N@5 & N@10 \\
\midrule
Caser    & 0.5217 & 0.6872 & 0.3571 & 0.4107  & 0.3095 & 0.4762 & 0.1993 & 0.2530  &  0.4620 & 0.5865 &  0.3435  & 0.3838   \\
GRU4Rec  & 0.5101 & 0.6723 & 0.3451 & 0.3976  & 0.3295 & 0.4856 & 0.2164 & 0.2670   & 0.4699 & 0.5994 & 0.3487 & 0.3906  \\
SASRec   & 0.5186 & 0.6829 & 0.3712 & 0.4242  & 0.4027 & 0.5439 & 0.2751 & 0.3210   & 0.4805 & 0.6083 & 0.3587 & 0.4000  \\
TiSASRec & 0.5313 & 0.6882 & 0.3812 & 0.4322  & 0.4014 & 0.5433 & 0.2745 & 0.3209  & 0.5114 & 0.6329 & 0.3860 & 0.4253  \\
\cmidrule(lr){1-1} \cmidrule(lr){2-5} \cmidrule(lr){6-9} \cmidrule(lr){10-13}
RCF      & 0.5101 & 0.6660 & 0.3635 & 0.4137  & 0.4145 & 0.5696 & 0.2911 & 0.3413  & 0.5790 & 0.7004 & 0.4475 & 0.4868  \\
RCF\textsubscript{\textit{LRD}} & 0.5398$^{\ddagger}$ & 0.6882$^{\ddagger}$ & 0.3886$^{\ddagger}$ & 0.4365$^{\ddagger}$  & 0.4381$^{\ddagger}$ & 0.5761$^{\ddagger}$ & 0.3127$^{\ddagger}$ & 0.3573$^{\ddagger}$  & 0.5828$^{\dagger}$ & 0.7035$^{\dagger}$ & 0.4510$^{\dagger}$  & 0.4901$^{\dagger}$  \\
Impro.   & +5.82\% & +3.33\% & +6.91\% & +5.51\%  & +5.69\% & +1.14\% & +7.42\% & +4.69\%  & +0.66\% & +0.44\% & +0.78\% & +0.68\% \\
\cmidrule(lr){1-1} \cmidrule(lr){2-5} \cmidrule(lr){6-9} \cmidrule(lr){10-13}
KDA      & 0.5748 & 0.7381 & 0.4182 & 0.4711  & 0.4453 & 0.6145 & 0.3127 & 0.3676  & 0.6008 & 0.7194 & 0.4665 & 0.5049  \\
KDA\textsubscript{\textit{LRD}} & \textbf{0.6066}$^{\ddagger}$ & \textbf{0.7434}$^{\ddagger}$ & \textbf{0.4420}$^{\ddagger}$  & \textbf{0.4867}$^{\ddagger}$  & \textbf{0.4826}$^{\ddagger}$ & \textbf{0.6302}$^{\ddagger}$ & \textbf{0.3403}$^{\ddagger}$ & \textbf{0.3881}$^{\ddagger}$  & \textbf{0.6111}$^{\ddagger}$ & \textbf{0.7295}$^{\ddagger}$ & \textbf{0.4760}$^{\ddagger}$  & \textbf{0.5143}$^{\ddagger}$  \\
Impro.   & +5.53\% & +0.72\% & +5.69\% & +3.31\%  & +8.38\% & +2.55\% & +8.83\% & +5.58\%  & +1.71\% & +1.40\% & +2.04\% & +1.86\% \\
\bottomrule
\end{tabular}
\end{table*}

\subsubsection{Baselines}
We selected multiple sequential recommendation models as baselines for our experiments:

\begin{itemize}
    \item \textbf{Caser}~\cite{tang2018personalized} adopts convolutional filters to capture sequential patterns of embed user sequences.
    \item \textbf{GRU4Rec}~\cite{hidasi2015session} utilizes Gated Recurrent Units (GRU) to model user representation by capturing patterns in the historical interaction sequence.
    \item \textbf{SASRec}~\cite{kang2018self} incorporates self-attention mechanisms to aggregate the historical item representations to obtain the user representation.
    \item \textbf{TiSASRec}~\cite{li2020time} further considers time intervals between historical interactions based on the SASRec.
    \item \textbf{RCF}~\cite{xin2019relational} adopts a two-level attention network to integrate item relations into the modeling of the user sequence representation.
    \item \textbf{KDA}~\cite{wang2020toward} incorporates a Fourier-based temporal evolution module to capture the dynamic changes in item relations over time.
\end{itemize}

Among these methods, \textbf{Caser}, \textbf{GRU4Rec}, \textbf{SASRec}, and \textbf{TiSASRec} belong to item-based collaborative filtering models, while \textbf{RCF} and \textbf{KDA} are models that explicitly incorporate item relations. Our proposed \textbf{LRD} method can discover latent item relations beneficial for recommendations to enhance existing relation-aware sequential recommendation models. Therefore, we compare the performance of \textbf{LRD}-enhanced \textbf{RCF} and \textbf{KDA}, i.e., \textbf{RCF}\textsubscript{\textit{\textbf{LRD}}} and \textbf{KDA}\textsubscript{\textit{\textbf{LRD}}}, with the baselines above.

\subsubsection{Evaluation Metircs}
We use two evaluation metrics, HR@K and nDCG@K, to evaluate the performance of the models, where K is set to 5 and 10. We adopt the leave-one-out method to construct the dataset. Specifically, for a user's interaction history sequence, we use the last item for testing, the second-to-last item for validation, and the remaining items for training. When predicting the next item, following previous work~\cite{wang2020toward}, we rank the ground-truth next item against 99 randomly sampled negative items. We report the average metrics over five runs with different random seeds.
\subsubsection{Implementation Details}
We implement LRD and baseline models with the ReChorus\footnote{https://github.com/THUwangcy/ReChorus/tree/master} library. We harness a widely used LLM, i.e., GPT-3\footnote{We use a version tailed for text embedding, i.e., text-embedding-ada-002}, to obtain language knowledge representations of items. The max length of the history sequence is set to 20. For all models, we use the Adam optimizer and carefully search for hyperparameters, with a batch size of 256 and embedding dimension of 64. The early stop is adopted if the nDCG@5 does not improve for 10 epochs. 
We tune the learning rate in \{1e-2, 1e-3, 1e-4\} and the l2-normalization coefficients in \{1e-4, 1e-5, 1e-6, 0\}. The coefficients of the knowledge graph embedding task and the latent relation discovery task are tuned within \{0.1, 1, 5, 10\}. The number of latent relations to discover is tuned between \{5, 6, 7, 8, 9, 10\}. The regularization coefficient of the relation extraction model of LRD is set to 0.1.
The code of our implementation is available at:\url{https://github.com/ysh-1998/LRD}.

\subsection{Performance Comparison (RQ1)}
We compare two LRD-enhanced relation-aware sequential recommendation models, namely RCF\textsubscript{LRD} and KDA\textsubscript{LRD}, with baseline methods. The overall experimental results are presented in the Tabel~\ref{tab: overall}. 

For baseline methods, several observations can be made. Firstly, traditional sequential recommendation models rely on implicit preferences in user interaction history to predict the next item. Thus, they heavily depend on rich interaction data. On the relatively dense MovieLens dataset, the performance gap between traditional methods and relation-aware methods is not significant. However, on the highly sparse Amazon datasets, traditional methods perform significantly worse than relation-aware methods. Secondly, by incorporating item relations into the sequential recommendation, RCF and KDA achieve significantly better performance than non-relation-aware sequential recommendation models. This indicates that explicitly modeling relations between historical items and the target item contributes to the user sequence representation modeling by capturing sufficient user preference. Thirdly, TiSASRec and KDA achieve significant performance improvement by incorporating modules that model temporal information based on SASRec and RCF, respectively. This demonstrates that considering temporal factors in modeling user interaction sequences can effectively enhance model performance.

The RCF\textsubscript{LRD} and KDA\textsubscript{LRD} achieve significant performance improvements over the vanilla models and outperform almost all non-relation-aware methods. We attribute these improvements to the fact that previous relation-aware sequential recommendation methods relied on predefined relations, limiting the model's ability to capture diverse item relations. Especially on datasets with fewer predefined relations, such as MovieLens, the performance of RCF is not significantly better than traditional methods. However, leveraging our proposed LRD method effectively improves the capability of relation-aware sequential recommendation models by capturing latent item relations. Moreover, the performance improvement of KDA\textsubscript{LRD} suggests that the discovered latent relations by LRD also exhibit temporal evolution characteristics.
\begin{figure}[t] 
  \centering
  \includegraphics[width=\linewidth]{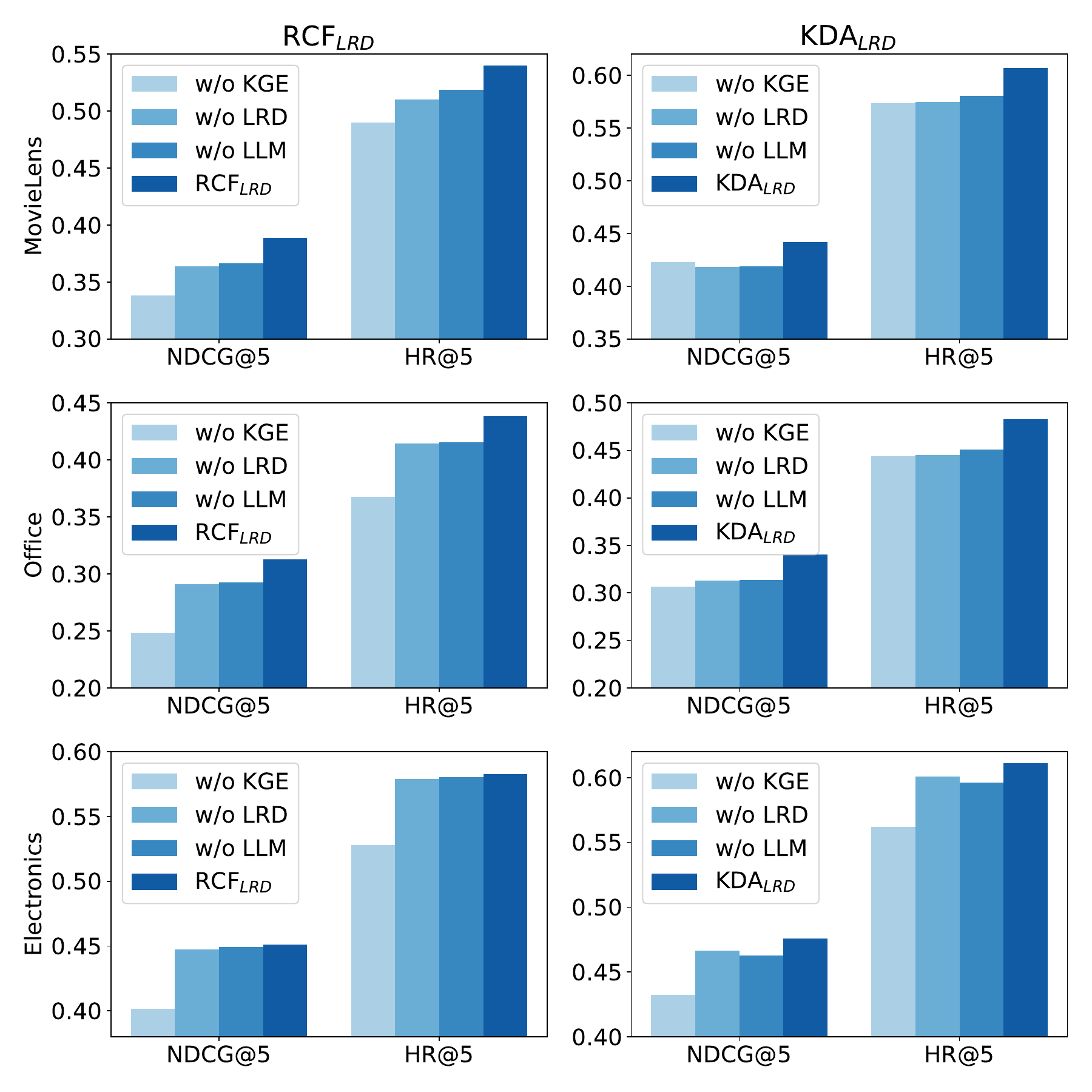}
  \caption{Ablation Study on variant models}
\label{fig: ablation}
\end{figure}

\subsection{Ablation study (RQ2)}
In this section, we investigate the contributions of each component of our proposed LDR-based relation-aware sequential recommendation model to the final recommendation performance. For this purpose, we design two variant models: (1) w/o LLM, where item ID representations replace the language knowledge representation obtained through LLM to investigate whether language knowledge is crucial in the relation discovery process. (2) w/o KGE, which removes the knowledge graph embedding task, meaning the model cannot optimize the representation of predefined relations utilizing supervised signals in the knowledge graph. The vanilla model without the LRD can also be seen as a variant model, i.e., w/o LRD.

Figure~\ref{fig: ablation} presents the comparative results of the variant models, revealing several findings. Firstly, removing the LLM for obtaining item language knowledge representations leads to a noticeable performance decline. This indicates that ID representations lacking rich semantic information have limited effectiveness in the process of discovering latent relations, emphasizing the crucial role of LLM's rich world knowledge. Secondly, the variant model without the KGE task shows a significant performance decrease, highlighting the necessity of maintaining the modeling of predefined relations while utilizing latent relations.

\subsection{Latent Relation Analyses (RQ3)}
To further validate the effectiveness and reliability of the latent item relations discovered by LRD, we perform additional analysis experiments. 

\subsubsection{Relation Embeddings} \label{sec: rel_emb}
\begin{figure}[t] 
  \centering
  \includegraphics[width=0.95\linewidth]{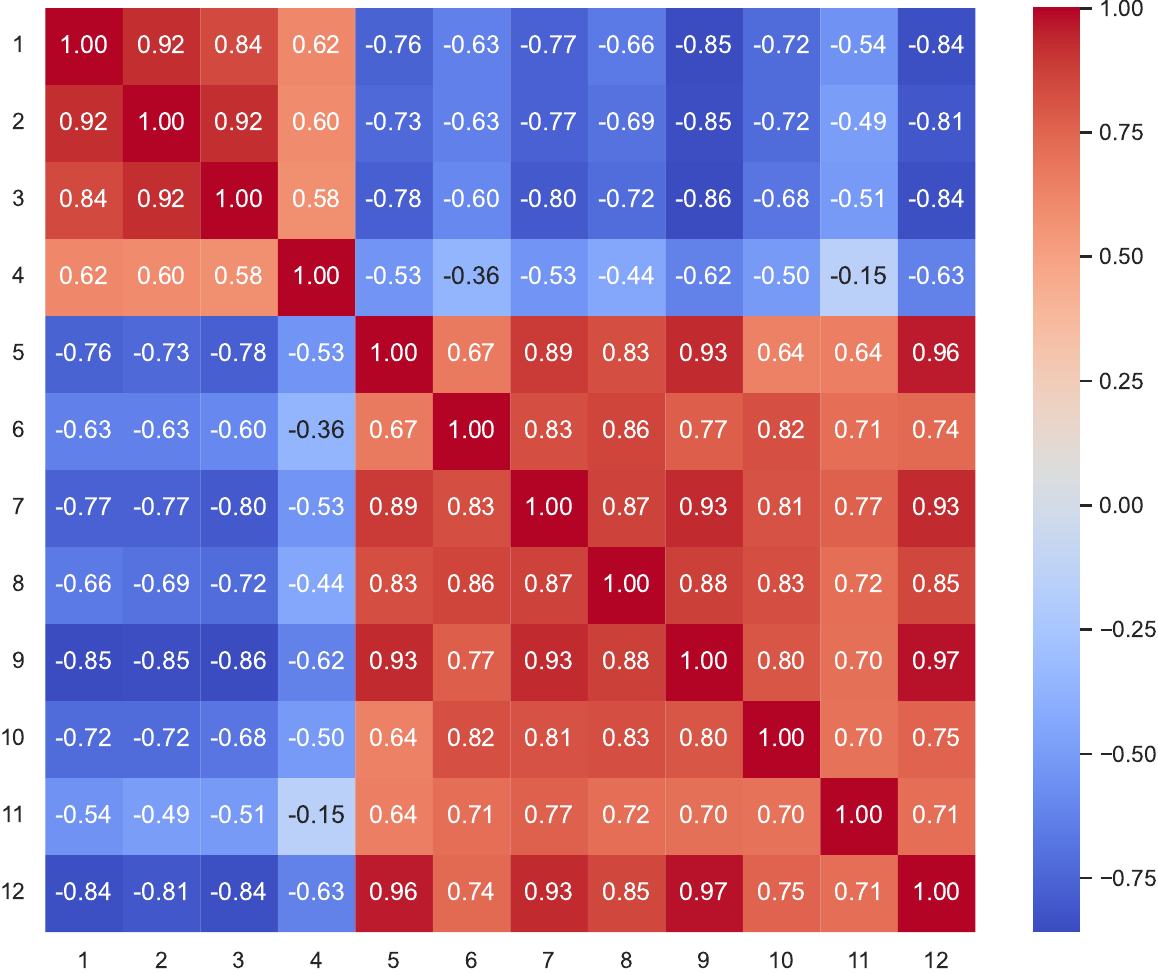}
  \caption{The pair-wise cosine similarity of the relation embeddings in KDA$_{LCD}$ on Office dataset. The labels on the horizontal and vertical axes denote the relation IDs, where 1-4 are predefined relations and 5-12 are latent relations learned by LRD.}
\label{fig: heatmap}
\end{figure}

To investigate the distribution of the learned relation embeddings, we select the model with the best performance on the Office dataset, i.e., KDA$_{LCD}$, and calculate the pair-wise cosine similarity of the relation embeddings. This model has 12 relations, including predefined relations (1-4) and latent relations (5-12). The similarity matrix is shown in Figure~\ref{fig: heatmap}. We can find that there is a notable difference between the embeddings of predefined relations and latent relations. This is reasonable as the learning of predefined relation embeddings primarily relies on supervised signals from the knowledge graph, while the learning of latent relation embeddings depends on the latent relation discovery task. It is worth noting that the relation extraction model of LRD also predicts predefined relations for the reconstruction of relational items. This indicates that the joint optimization of the latent relation discovery task and the knowledge graph embedding task helps the model effectively distinguish predefined relations from latent relations during the learning process. 


\subsubsection{Latent Relational Items} \label{sec: rel_case}
\begin{figure}[t] 
  \centering
  \includegraphics[width=\linewidth]{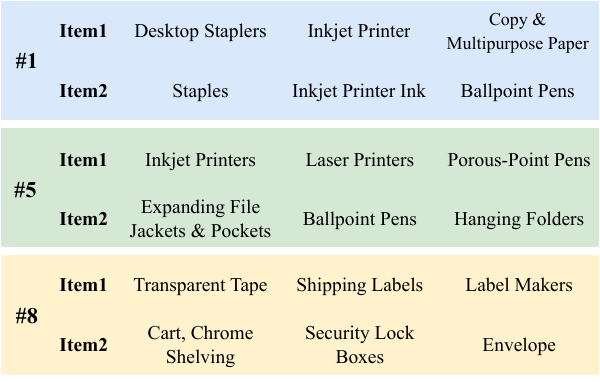}
  \caption{Descriptions of representative item pairs belong to relation \#1, \#5 and \#8 of KDA$_{LCD}$ on Office dataset, where relation \#1 is ``also buy”, relation \#5 and \#8 are learned latent relations.}
\label{fig: rel_example}
\end{figure}

To provide a clearer illustration of the learned item relations, we present representative item pairs of several relation types in Figure~\ref{fig: rel_example}. These pairs are obtained by sorting the model's scores for these item pairs on these relations, i.e., the selected item pairs are samples with higher scores under each relation. To facilitate readability, we have simplified the item descriptions. 
Several observations can be made from Figure~\ref{fig: rel_example}. 
Firstly, items in relation \#1 generally exhibit complementary functionalities and are frequently purchased together. This indicates that the model effectively learns predefined relations from the knowledge graph, i.e., the ``also buy” relation. Secondly, the model groups item pairs with common characteristics into the same latent relation type. These relations are more complex than predefined relations, possibly involving multi-hop connections and demonstrating characteristics tailored to specific scenarios and tasks. The relation \#5 reflects the relations between items in document processing, editing, and storage scenarios. Items like ``Inkjet Printers” and ``Expanding File
Jackets \& Pockets”, though not directly related, can establish a two-hop relation through ``printing paper”, completing a document processing procedure. Relation \#8 focuses on tasks related to transportation. Items like ``Transparent Tape” and ``Cart, Chrome Shelving” are related to the task of packing and transporting goods.

These examples further demonstrate the reliability of the latent relations discovered by LRD and confirm the conclusion drawn in Section~\ref{sec: rel_emb} that latent relations differ significantly from predefined relations. Leveraging these more complex and diverse relations, the LRD-based relation-aware sequential recommendation model effectively captures more intricate user preferences, resulting in improved recommendation performance.

\subsubsection{Case Study}
\begin{figure}[t] 
  \centering
  \includegraphics[width=\linewidth]{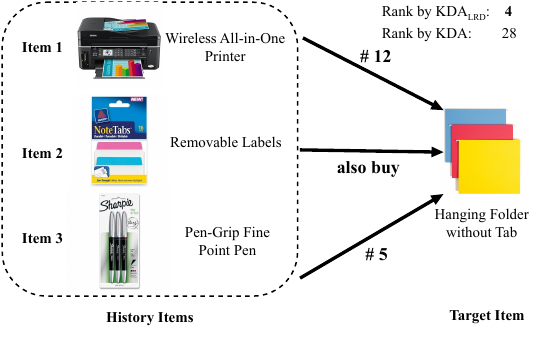}
  \caption{The interaction history of user A3N4VTNFPMTHEF on Office dataset. KDA$_{LCD}$ ranks the target at a high position leveraging the learned latent relation between historical items and target item, i.e., relation \#5 and \#12, which outperforms KDA solely utilizing predefined relation, i.e. ``also buy”.}
\label{fig: latent_case}
\end{figure}

To further demonstrate how the model leverages latent relations to achieve improved recommendation performance, we present a case in Figure~\ref{fig: latent_case}. In this case, we record the triplet scores between each historical item and the target item on all relations. The relation with the highest score is considered as the predicted relation between items. As shown in Figure~\ref{fig: latent_case}, only one predefined relation exists between historical items and the target item, i.e., ``also buy.” Relying solely on this predefined relation, KDA fails to rank the target item at a high position. While LRD-based model KDA\textsubscript{\textit{LRD}}, identifies latent relations \#5 and \#12 between items 1 and 3, and the target item, respectively. Considering the findings from Sections~\ref{sec: rel_emb} and Section~\ref{sec: rel_case}, latent relations \#5 and \#12 exhibit similar embeddings and both reflect item relations in the document processing scenario. Specifically, in this case, the printout from the ``Wireless All-in-One
Printer", the document recorded with the ``Pen-Grip Fine
Point Pen", and both can be stored in the ``Folder". Benefiting from the discovered latent relations between historical items and the target item by LRD, KDA\textsubscript{\textit{LRD}} ranks the target item in the 4th position, significantly outperforming the vanilla model KDA.

\subsection{Hyper-parameter Sensitivity}
\begin{figure}[t] 
  \centering
  \includegraphics[width=\linewidth]{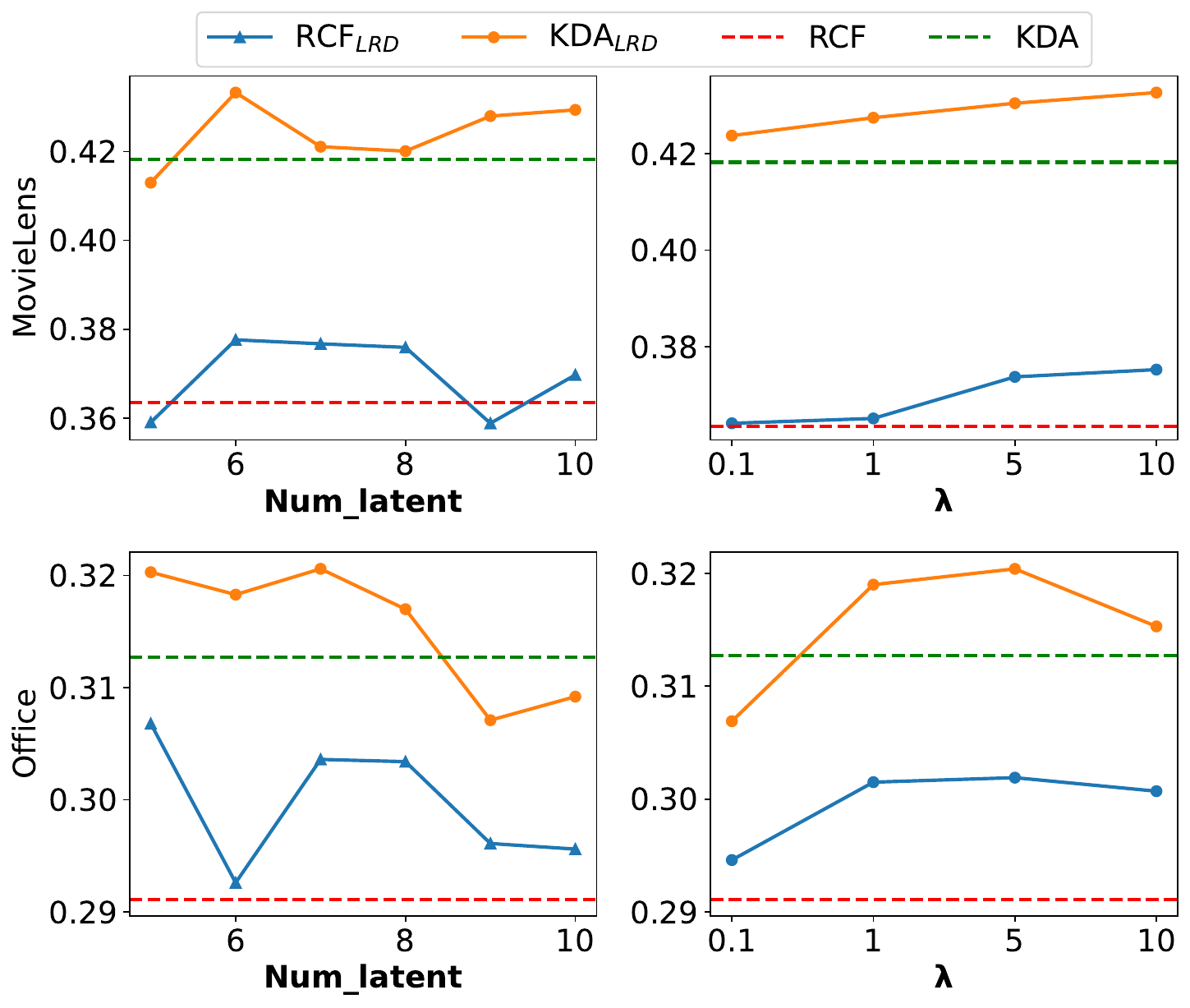}
  \caption{nDCG@5 comparison w.r.t. the number of latent relations and the coefficient of the latent relation discovery task, i.e., $\lambda$.}
\label{fig: hyper}
\end{figure}

In this section, we perform a sensitivity analysis on two crucial hyperparameters of LRD: the number of latent relations the model aims to discover, i.e., Num\_latent, and the coefficient of the latent relation discovery task in the joint optimization, i.e., $\lambda$. We aim to analyze the effect of hyperparameter selection on model performance. Firstly, we examine the sensitivity of the model performance to different values of num\_latent. Note that to isolate the impact of the two hyperparameters, we present the average performance under the same num\_latent value across all $\lambda$ values. Figure~\ref{fig: hyper} illustrates the model performance under various num\_latent values and several observations can be made. Firstly, the model exhibits significant performance differences under different num\_latent values, indicating that both learning an insufficient number of latent relations and learning redundant or useless relations impair the model performance. Secondly, the overall trends of performance changes with num\_latent are generally consistent across different models on the same dataset. While the optimal num\_latent value varies for different datasets. Specifically, the optimal value for MovieLens is 6, while for the Office dataset is 5. This aligns with the assumption that there are differences in item relations across diverse recommendation scenarios.
Next, we analyze the hyperparameter $\lambda$. The performance shown in Figure~\ref{fig: hyper} represents the average under the same $\lambda$ value across all num\_latent values. We observe a stable improvement in model performance until the optimal value is reached, which is generally consistent on the same dataset for two models. That demonstrates the effectiveness of the latent relation discovery task.

\section{Related Work}
\subsection{Sequential Recommendation}
In the literature of recommender systems, sequential recommendation is a widely researched task, aiming to recommend items that interest users based on their historical interactions~\cite{wang2019sequential}. Earlier works adopt the Markov chain to model item transition relations in the interaction history of users~\cite{rendle2010factorizing,he2016fusing}. In recent years, with the development of deep learning methods, various deep neural networks have been proposed to capture user preferences in historical interaction sequences. It includes Recurrent Neural Networks (e.g., GRU~\cite{hidasi2015session}, LSTM~\cite{wu2017recurrent} and HRNN~\cite{quadrana2017personalizing}), Convolutional Neural Networks~\cite{tang2018personalized,yuan2018simple}, Attention-based Network~\cite{kang2018self,sun2019bert4rec,li2017neural}, and Graph Neural Networks~\cite{wu2019session,chang2021sequential}. The core idea of these methods is to capture item-based collaborative filtering information in historical interactions, overlooking explicit relations between items, which is crucial for understanding user behavior and extracting user preferences more effectively. The proposed approach in this paper can effectively uncover relations among items, thus providing a more efficient modeling of user preferences.

\subsection{Relation-aware Recommendation}
In contrast to traditional recommendation methods that consider only item-based collaborative filtering similarity, relation-aware recommendation methods explicitly incorporate relations between items into the recommendation model. One line of methods constructs a knowledge graph containing relations between items and relations between items and attributes~\cite{zhang2016collaborative,ai2018learning,wang2018dkn,cao2019unifying,wang2019multi,ma2019jointly}. These methods enhance item representations through knowledge graph embedding tasks. CKE~\cite{zhang2016collaborative} constructs a knowledge graph containing items and attributes and optimizes graph embedding tasks and recommendation tasks jointly. CFKG~\cite{ai2018learning} further incorporates users into the knowledge graph. It defines a special ``purchase'' relation as the proxy of the interactions between users and items to transform the recommendation task into the knowledge graph embedding task.
Another research direction focuses on sequential recommendation scenarios and explicitly models relations between historical items and the target item~\cite{xin2019relational,wang2020make,wang2020toward}. RCF~\cite{xin2019relational} proposes a two-level attention framework to compute the user's attention to relation types and the relation intensity between historical items and the target item separately. Furthermore, considering the evolution of item relations over time, KDA~\cite{wang2020toward} proposes a Fourier-based temporal evolution module to incorporate time information into the modeling of relational items. Nevertheless, existing methods rely on manually predefined relations and suffer from relation sparsity and item sparsity issues. In this paper, we propose to uncover latent relations among items, enabling the model to adapt to diverse and complex recommendation scenarios.

\subsection{Large Language Model for Recommendation}
Recently, the emergence of Large Language Models (LLMs)~\cite{brown2020language,zhang2022opt,ouyang2022training,touvron2023llama,openai2023gpt4} has revolutionized the field of natural language processing. The performance of various natural language processing tasks has significantly improved with the reasoning abilities and world knowledge of LLMs~\cite{kamalloo2023evaluating,zhang2023benchmarking}. The application of LLMs to recommendation tasks has been widely researched in two main research directions~\cite{fan2023recommender}. One line of work leverages LLMs' reasoning capabilities and knowledge to solve various recommendation tasks.
Liu et al.~\cite{liu2023chatgpt} utilize ChatGPT~\cite{openai2023gpt4} to perform five recommendation tasks in zero-shot and few-shot settings, achieving impressive results in explanation generation tasks while performing poorly in the direct recommendation and sequential recommendation tasks. P5~\cite{geng2023recommendation} transform various recommendation tasks into natural language format and feed into T5~\cite{raffel2023exploring}. Then the model is fine-tuned with the language modeling training objective and exhibits performance improvement on five recommendation tasks.
Another line of research investigates combining recommendation models with LLMs. ChatRec~\cite{gao2023chatrec} proposes to first recall candidate items from a vast number of items utilizing traditional recommendation models and further adopts an LLM to re-rank candidate items, demonstrating the potential of using LLM as rerankers. TALLRec~\cite{Bao_2023} constructs instruction-tuning samples based on the user rating data and fine-tune LLaMA~\cite{touvron2023llama} with parameter-efficient-tuning, i.e., LoRA~\cite{hu2021lora}, achieving promising performance on few-shot recommendation scenario. InstructRec~\cite{zhang2023recommendation} organizes diverse instruction samples, including user interaction data and comments, to perform instruct-tuning on FLANT5~\cite{chung2022scaling}. The trained model contributes to the recommendation model as a reranker. ONCE~\cite{liu2023once} utilizes ChatGPT as a data augmenter to acquire knowledge-enhanced representations of users and news, improving the performance of the news recommendation.
However, existing recommendation methods with LLMs have significant limitations in terms of performance and effectiveness. In this paper, we leverage the rich world knowledge of the LLM to acquire knowledge representations of items, which is used to discover item relations that contribute to recommendations. We fully utilize the knowledge of the LLM and ensure the efficiency of the overall framework.

\section{Conclusion}
In this paper, we propose a novel method for discovering latent item relations based on the Large Language Model (LLM), namely LRD. Leveraging the rich world knowledge of LLM and a self-supervised learning approach, LRD effectively extracts latent item relations. We jointly optimize LRD with existing relation-aware sequential recommender systems. On the one hand, the latent relations discovered by LRD provide more sophisticated item associations, contributing to the sufficient modeling of intricate user preference. On the other hand, the supervision signals from user interactions guide the relation discovery process effectively. Experimental results on multiple public datasets demonstrate that LRD significantly improves the performance of existing relation-aware sequential recommendation methods. Further analyses demonstrate the reliability of the latent relations. Note that the crucial LLM in our method is not meticulously selected in the current implementation. It leaves us a future work of exploring the performance of more advanced LLMs in our method.


\balance
\bibliographystyle{ACM-Reference-Format}
\bibliography{sample}










\end{document}